\documentstyle[aps,prb,multicol,graphicx,epsfig]{revtex}

\begin{document}
\title{Crosstalk between nanotube devices: contact and channel effects}
\author{Fran\c{c}ois L\'{e}onard}
\address{Sandia National Laboratories, Livermore, CA 94551}
\date{\today }
\maketitle
\draft

\begin{abstract}
At reduced dimensionality, Coulomb interactions play a crucial role in
determining device properties. While such interactions within the same
carbon nanotube have been shown to have unexpected properties, device
integration and multi-nanotube devices require the consideration of
inter-nanotube interactions. We present calculations of the characteristics
of planar carbon nanotube transistors including interactions between
semiconducting nanotubes and between semiconducting and metallic nanotubes.
The results indicate that inter-tube interactions affect both the channel
behavior and the contacts. For long channel devices, a separation of the
order of the gate oxide thickness is necessary to eliminate inter-nanotube
effects. Because of an exponential dependence of this length scale on
dielectric constant, very high device densities are possible by using high-$%
\kappa $ dielectrics and embedded contacts.
\end{abstract}

\begin{multicols}{2}

\section{Introduction}

Much recent experimental and theoretical work has focused on the electronic
transport properties of nanowires, motivated by promises of novel electronic
devices and the basic scientific challenges that they present. One of the
key findings in nanowire devices is that the Coulomb interaction plays a
crucial role in determining device properties. For example, electrostatics
in carbon nanotubes (NTs) leads to unexpected behavior of intra-tube {\it p-n%
} junctions\cite{leonard1}, of intra-tube Schottky junctions\cite{odintsov},
and of Fermi level pinning at contacts between NTs and metals\cite{leonard2}%
. The importance of {\it intra}-tube Coulomb interactions naturally leads to
the question of how {\it inter}-tube interactions might influence device
behavior. Answering this question is gaining increasing importance as
experimental devices based on multiple nanotubes are becoming more common%
\cite{star,snow,li,dimaki} [an example is shown in Fig. 1(a)], and also to
address the question of device packing density.

In this paper, we present self-consistent calculations based on the
non-equilibrium Green's function technique for planar carbon nanotube
transistors containing multiple parallel NTs. By analyzing the variations of
the transistor characteristics with NT separation, we establish a length
scale below which inter-tube interactions become important. While for small
channel lengths this length scale depends on the channel length, for long
channel devices it becomes independent of the channel length and is
essentially determined by the gate oxide thickness. Importantly, this length
scale depends exponentially on the dielectric constant of the medium
surrounding the NTs, and can be substantially reduced by embedding the NTs
in a high-$\kappa $ dielectric in the channel and using embedded contacts.

\section{Computational approach}

Figure 1(b-e) shows the device under consideration: infinitely long
single-wall zigzag NTs laying on metal electrodes at their two ends and on a
dielectric in the channel region. A planar gate 10 nm below the dielectric
surface controls the device behavior. We take a spacing of 0.3 nm between

\begin{figure}[h]
\psfig{file=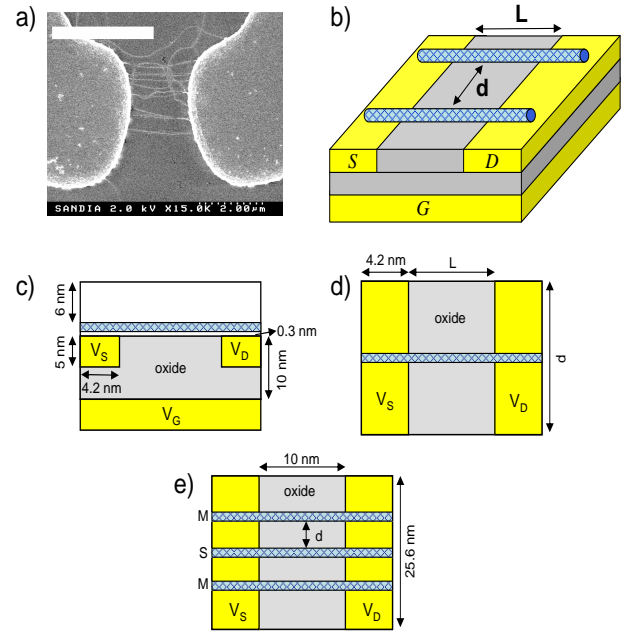,height=250pt,width=240pt}
\caption{Panel (a) shows a SEM image of a carbon nanotube device containing
multiple nanotubes. Panel (b) shows a sketch of the device used in our
calculations, with the distance $L$ indicating the channel length and $d$
the NT separation. Panel (c) shows a side view of the device, while panels
(d) and (e) show top views of the unit cells used to study
semiconducting-semiconducting and semiconducting-metallic interactions,
respectively.}
\end{figure}

the NT and the metal or dielectric, and assume that the NT cross-section
remains circular, a good approximation for the NTs considered here\cite%
{deformation}. The electronic properties of the NT are described using a
tight-binding approach with one $\pi $ orbital per carbon atom, and a
coupling $\gamma =2.5$ eV between nearest-neighbor atoms. Semiconducting
tubes are (17,0) NTs (band gap of 0.55 eV, diameter of 1.33 nm) while
metallic tubes are (18,0) NTs (diameter of 1.41 nm). We take the NT midgap
as the energy reference level, and use a metal workfunction of 5.5 eV,
putting the metal Fermi level 1 eV below the NT midgap. The gate insulator
dielectric constant is that of SiO$_{2}.$

To apply the non-equilibrium Green's function formalism\cite%
{datta,leonardiccn} to this system, we divide the NT in principle layers,
with each layer corresponding to a ring of the zigzag NT. The main quantity
of interest is the Green's function $G^{R}$, from which one can obtain the
transmission probability $T(E)$ and the zero bias conductance

\begin{equation}
C=\frac{2e\gamma }{\hbar }\int T(E)\left[ -\frac{\partial f(E)}{\partial E}%
\right] dE
\end{equation}%
where $f$ is the Fermi function. $G^{R}$ is calculated by solving the matrix
equation

\begin{equation}
G^{R}=\left[ \left( E-eU\right) I-H_{0}-\Sigma ^{R}\right] ^{-1}\text{,}
\label{Gr}
\end{equation}%
where $H_{0}$ is the tight-binding Hamiltonian for the isolated NT and $U$
is the electrostatic potential evaluated at the position of each layer. The
functions $\Sigma ^{R}$ and $\Sigma ^{<}$ represent the interaction of the
scattering region with the semi-infinite NT leads. In our tight-binding
representation, the Hamiltonian matrix elements for layer $l$ are $%
H_{0}^{2l,2l-1}=H_{0}^{2l-1,2l}=2\gamma \cos \left( \frac{\pi J}{M}\right) $%
, $H_{0}^{2l,2l+1}=H_{0}^{2l+1,2l}=\gamma $ where $M$ is the number of atoms
around a NT ring ( $M=17$ or $18$ here) and $J=1,...,M$ labels each of the
NT bands. In this representation, we assume that the electrostatic potential
on every atom of a ring is the same, a good approximation in this case since 
$U$ varies slowly over the NT diameter.

The electrostatic potential is calculated by solving Poisson's equation in
three-dimensional coordinates on a variable grid with the source charge on
the NT, and with boundary conditions at the source, drain, and gate
surfaces, and at the vacuum/dielectric interface. This is done by combining
a finite difference approach in the directions parallel to the NT and
perpendicular to the substrate with a fast fourier transform in the
direction perpendicular to the NTs. Once the three-dimensional electrostatic
potential is obtained, the value for $U$ on each ring is taken as the
average of the electrostatic potentials on the points on the NT nearest and
furthest from the substrate, at the position of each ring.

To obtain the charge density, we note that our tight-binding technique
provides the total charge on each layer of the NT, which needs to be
spatially distributed. We assume a uniform distribution of the charge in the
azimuthal direction, and spatially distribute the total charge in the radial
and axial directions with a Gaussian smearing function. The
three-dimensional charge density is then given by 
\begin{equation}
\sigma (r,z,\phi )=-\sum_{l}g(z-z_{l},r-R)\frac{e}{\pi }\int dE%
\mathop{\rm Im}%
G_{ll}^{R}  \label{charge}
\end{equation}%
where $g(z-z_{l},r-R)=\left( 4\pi ^{2}R\sigma _{z}\sigma _{r}\right)
^{-1}\exp \left[ -(z-z_{l})^{2}/2\sigma _{z}^{2}\right] \\
\exp \left[
-(r-R)^{2}/2\sigma _{r}^{2}\right] $ with $R$ the tube radius, $z_{l}$ the
position of ring $l$, and $\sigma _{z}$ and $\sigma _{r}$ the smearing
lengths in the axial and radial directions respectively (this expression for 
$g$ is valid when $R\gg \sigma _{r}$, and we used values of $\sigma
_{z}=0.14 $ nm and $\sigma _{r}=0.06$ nm). Our overall procedure is to solve
the coupled set of nonlinear Eqs $\left( \ref{Gr}-\ref{charge}\right) $
self-consistently for a given gate-source voltage at zero drain-source bias.
Figure 1(e) shows the unit cells for our calculations to study
semiconducting-semiconducting and semiconducting-metallic interactions.

\begin{figure}[h]
\psfig{file=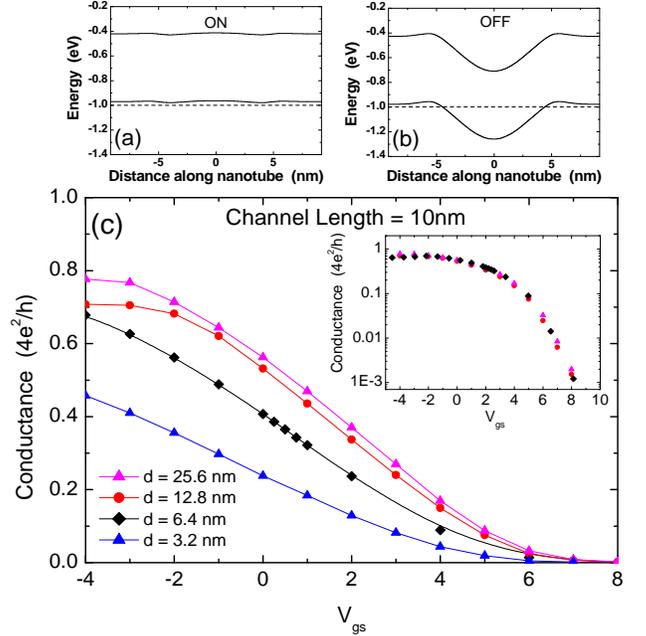,height=250pt,width=240pt}
\caption{Panels (a) and (b) show band diagrams for $d=25.6$ nm at
gate-source voltages of -4 and 8 volts, respectively. Panel (c) shows the
conductance of a (17,0) NT transistor as a function of gate-source voltage,
for different NT separations. The inset shows collapse of the data after
rescaling and shifting of the gate voltage.}
\end{figure}

\section{Results and discussion}

We first discuss interactions between semiconducting NTs. The conductance of
the NT transistor as a function of the gate-source voltage is plotted in
Fig. 2 for different values of the separation between the NTs. Clearly, as
the NT separation is reduced, the NTFET characteristics degrade; in the
range of gate voltages studied, the current cannot be completely turned on
for the smallest tube separation, while the behavior in the OFF state is
essentially independent of tube separation. This behavior arises because of
the charge induced on the NTs by the gate voltage, and by interactions
between NTs at the contact. As shown in Fig. 2a, the OFF state corresponds
to the Fermi level in the middle of the NT bandgap in the channel, with
little charge on the NTs, so Coulomb interactions in the channel between
neighboring NTs are negligible. The ON state however consists of making the
NT {\it p}-type in the channel by raising the valence band above the Fermi
level with the gate voltage (Fig. 2b); this creates a large positive charge
on the NT which interacts with the charge on neighboring NTs, and lowers the
electron energy. This pushes the bands down, and reduces the conductance.

While interactions in the channel play an important role in determining
device behavior, NT-NT interactions also affects the contacts. Because of
the mismatch between the metal and NT workfunctions, contacts between NTs
and metals are governed by charge transfer between the metal and the NT,
effectively doping the NT {\it p}-type at the contact, with the metal Fermi
level just below the NT valence band edge in the contact\cite{leonard3}. As
the separation between NTs decreases, the Coulomb interactions between the
positive charge at the contacts leads to a lowering of the bands; at small
enough separations, the valence band in the contact can be pushed below the
metal Fermi level. Thus inter-tube interactions at the contacts can lead to
modifications of the NT/metal contacts from ohmic to Schottky. We find that
this effect becomes important for inter-tube separations below $5$ nm; the
effects can be seen in the bottom curve of Fig. 2, where the ON state
conductance saturates to a value much smaller than the single-tube limit
because of the Schottky barrier. These effects of inter-tube interactions on
contact properties can be mitigated by embedding the NTs in the metal
contact.

\begin{figure}[h]
\psfig{file=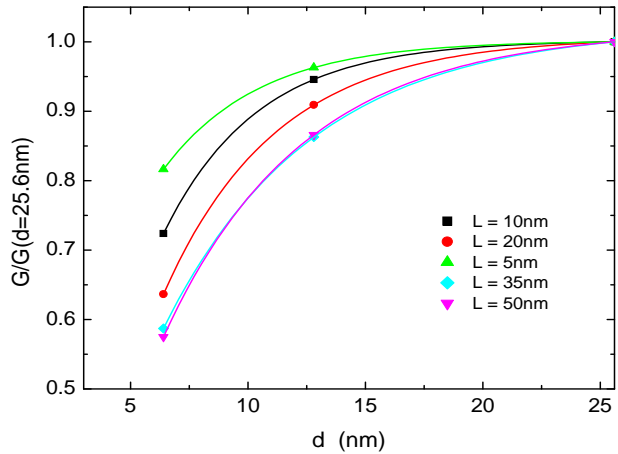,height=180pt,width=240pt}
\caption{Conductance at $V_{gs}=0$ versus NT separation, for different
channel lengths. Solid lines are fits of the form $A(1-e^{-d/\protect\lambda %
})$.}
\end{figure}

To quantify the role of Coulomb interactions {\it in the channel}, we
calculated the conductance versus gate voltage for devices with different
channel lengths; we plot in Fig. 3 the conductance at $V_{gs}=0$ versus tube
separation, for several values of the channel length (the smallest value of $%
d$ is large enough to avoid contact effects). The solid lines in the figure
are fits of the form $A\left( 1-e^{-L/\lambda }\right) $, from which we
extract the value of $d$ at which the conductance decreases by more than
10\% from its $d\rightarrow \infty $ value. This length scale is plotted in
Fig. 4 as a function of channel length, and delimits regions where
inter-tube interactions distort the single-tube behavior. Clearly, very
short channel devices can have very high packing densities, while long
channel devices are limited to tube separations of 15 nm. A notable aspect
of the results of Fig. 4 is that the value of $d$ is independent of the
channel length for large $L$. At first glance, one would expect that a
larger channel length leads to larger total charge on the NTs and thus
larger interaction energy. However, screening of the Coulomb interaction by
the planar gate leads to a different behavior, as we now discuss.

\begin{figure}[h]
\psfig{file=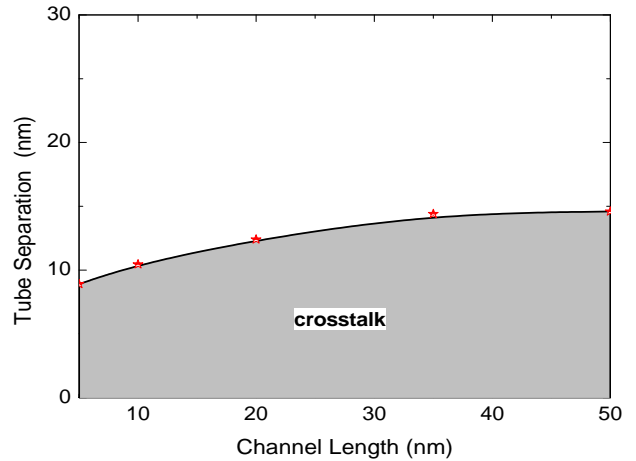,height=180pt,width=240pt}
\caption{Tube separation below which inter-tube interactions become
important. Note the saturation of the length scale as the channel length
increases.}
\end{figure}

To derive approximate analytical expressions for the Coulomb interactions
between NTs, we consider the simplified situation of Fig. 5a: a NT of length 
$L$ is at a distance $L_{g}$ above a metallic substrate held a potential $%
V_{g}$, and carries charge per unit length $\lambda $. We use an image
potential construction to calculate the electrostatic potential $V$ at a
distance $d$ from the charged tube, giving

\begin{equation}
V=V_{g}+\frac{\lambda }{4\pi \varepsilon }\ln \frac{\left( L+\sqrt{%
L^{2}+4d^{2}}\right) \left( -L+\sqrt{L^{2}+4d^{2}+16L_{g}^{2}}\right) }{%
\left( -L+\sqrt{L^{2}+4d^{2}}\right) \left( L+\sqrt{L^{2}+4d^{2}+16L_{g}^{2}}%
\right) }.  \label{V}
\end{equation}%
In the long channel limit $L\gg d,L_{g}$, Eq. (\ref{V}) becomes%
\begin{equation}
V=V_{g}+\frac{\lambda }{2\pi \varepsilon }\ln \frac{\sqrt{d^{2}+4L_{g}^{2}}}{%
d}.  \label{semisemi}
\end{equation}%
The important point here is that in this limit the potential shift is
independent of the channel length $L$. Taking a potential change of $0.1$
Volts as a criterion for the importance of inter-tube effects, we obtain a
separation%
\begin{equation}
d^{\ast }=\frac{2L_{g}}{\sqrt{e^{\frac{0.4\pi \kappa \varepsilon _{0}}{%
\lambda }}-1}}
\end{equation}%
below which inter-tube effects become important. Thus the gate oxide
thickness $L_{g}$ sets the length scale $d^{\ast }$. In the above
expression, $\kappa $ is the effective dielectric constant for the device
geometry. For a simple comparison with our simulations we take $\kappa $ as
the dielectric constant of SiO$_{2}$, and use the computed value of $\lambda
=1\times 10^{-3}$ e/C atom in the ON state. This gives $d^{\ast }\approx 10$
nm in good agreement with our numerical results. A key point is that the
length scale $d^{\ast }$ depends {\it exponentially} on the gate insulator
dielectric constant; thus, replacing SiO$_{2}$ with high-$\kappa $
dielectric materials (NTFETs have been fabricated with HfO$_{2}$\cite{javey1}%
, SrO$_{2}$\cite{javey2} and SrTiO$_{3}$\cite{kim}) allows a reduction of $%
d^{\ast }$ to extremely small values, thus permitting high device densities.

\begin{figure}[h]
\psfig{file=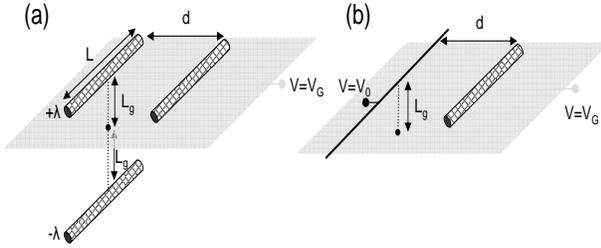,height=100pt,width=240pt}
\caption{Sketch of situations considered for analytical calculation of
interactions between (a) semiconducting nanotubes, (b) semiconducting and
metallic nanotubes.}
\end{figure}

The general behavior of interactions between semiconducting NTs can be
understood from further analysis of Eq. $\left( \ref{semisemi}\right) $.
Near threshold, the charge on the NT can be written as $\lambda =\lambda
_{0}\left( V_{g}-V_{th}\right) $ where $V_{th}$ is the threshold voltage.
Substitution in Eq. $\left( \ref{semisemi}\right) $ gives%
\begin{equation}
V=\left( 1+\frac{\lambda _{0}}{2\pi \varepsilon }\ln \frac{\sqrt{%
d^{2}+4L_{g}^{2}}}{d}\right) V_{g}-\frac{\lambda _{0}V_{th}}{2\pi
\varepsilon }\ln \frac{\sqrt{d^{2}+4L_{g}^{2}}}{d}.
\end{equation}%
Hence, the gate voltage is rescaled and shifted by the interaction between
NTs. This behavior is indicated in the inset of Fig. 2c, showing good
collapse of the data after appropriate rescaling and shifting of $V_{g}$.

The results discussed above focused on interactions between semiconducting
NTs. However, in multi-NT devices, there is often a mixture of
semiconducting and metallic nanotubes. A question therefore is how the
presence of nearby metallic NTs affects the properties of the semiconducting
NTs. To address this issue, we performed calculations for the geometry
depicted in Fig. 1(e). There, a semiconducting NT is separated from two
metallic NTs by a distance $d$. Figure 6 shows the calculated conductance vs
gate voltage dependence for a 10 nm channel device. (The conductance plotted
here is that of the semiconducting NT only. For single electrodes making
contact to all the NTs, the conductance of the metallic tubes would have to
be included to obtain the total device conductance.) The behavior in Fig. 6
is similar to that observed for interactions between semiconducting NTs,
i.e. a degradation of the characteristics with decreasing separation between
the NTs. However, because the charge on the metallic tubes is not strongly
modulated by the gate voltage, the physics is somewhat different; indeed,
the metallic NTs essentially consist of lines of constant potential. To
their effect on the semiconducting NTs we consider the situation of Fig. 5b,
where a line held at a constant potential $V=V_{0}$ is above a metallic
plane held at potential $V_{g}$. Solution of the Laplace equation for this
geometry gives the potential on the nearby NT as%
\begin{equation}
V=V_{0}\frac{L_{g}^{2}}{d^{2}+L_{g}^{2}}+V_{g}\frac{d^{2}}{d^{2}+L_{g}^{2}}.
\label{scaling}
\end{equation}%
Thus the presence of the metallic tubes shifts and rescales the gate voltage
seen by the semiconducting tube. For small separations between the NTs, the
potential on the semiconducting NT approaches that of the metallic NT, which
is larger than that of an isolated semiconducting NT. Thus, at the contact,
the Fermi level is pushed above the valence band edge, leading to a Schottky
barrier and saturation of the conductance to lower values. The behavior
expressed in Eq. (\ref{scaling}) can be verified by looking at the OFF state
behavior, and scaling and shifting the gate voltage appropriately; the inset
in Fig. 6 shows the resulting collapse of the data.

\begin{figure}[h]
\psfig{file=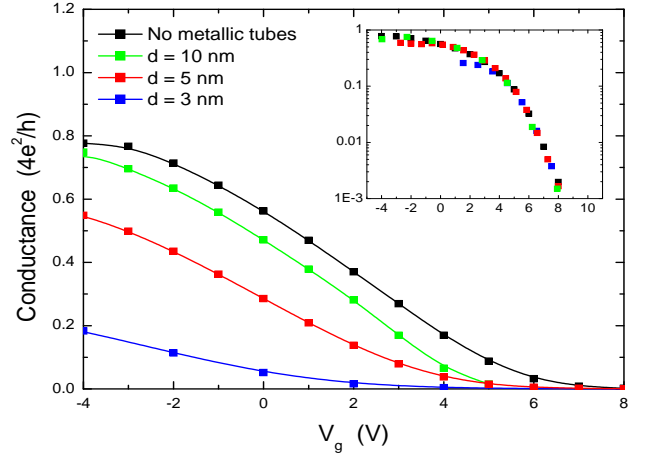,height=180pt,width=240pt}
\caption{Conductance of semiconducting nanotube versus gate voltage for a
channel length of 10 nm, as a function of separation from neighboring
metallic nanotubes.}
\end{figure}

\section{Conclusion}

In summary, we have shown that Coulomb interactions between carbon nanotubes
can have a strong influence on device behavior. Below a characteristic
length scale, interactions between NTs can significantly degrade device
behavior; this can be understood in terms of a rescaling and shift of the
applied gate voltage. For long channel devices, the tube separation above
which interactions between NTs is negligible becomes independent of the
channel length, and is set by the gate oxide thickness. This length scale
can be substantially reduced by using high-$\kappa $ dielectrics, due to an
exponential dependence on dielectric constant. While this paper focused on
the static device properties, we expect that interactions during current
flow (e.g. Coulomb drag) would also have intriguing properties. We hope that
this work will stimulate controlled experiments to further explore
interactions between nanotubes.

{\bf Acknowledgments}

Sandia is a multiprogram laboratory operated by Sandia Corporation, a
Lockheed Martin Company, for the United States Department of Energy under
contract DE-AC01-94-AL85000. We thank K. McDonald, F. Jones, and A. Talin
for providing the SEM image in Fig. 1.

\end{multicols}

\end{document}